\documentclass[letter]{aa}     

\usepackage{graphicx}
\usepackage{txfonts}
\usepackage{lipsum}
\usepackage{subcaption}         
\usepackage{lscape}             
\usepackage{placeins}           

\usepackage{natbib} 
\usepackage{float}
                                
\begin{document}

   \title{Beyond Bragg-Mirrors for Gravitational Wave Telescopes:}

   \subtitle{A Fabrication Tolerant Hybrid Metasurface-Bragg Mirror Design}

%
%
%

   \author{C. Kranhold\inst{1}\corrauth{christian.kranhold@uni-jena.de}        
        \and M. Gaedtke\inst{2, 3}\email{mika.gaedtke@tu-braunschweig.de}
        \and M. Walther\inst{1}\email{walther.markus@uni-jena.de}
        \and F. Eilenberger\inst{1, 4}\email{falk.eilenberger@iof.fraunhofer.de}
        \and S. Kroker\inst{2, 3, 5}\email{stefanie.kroker@tu-braunschweig.de}
        \and T. Siefke\inst{1, 4, 6}\email{thomas.siefke@uni-jena.de}
        }

   \institute{Friedrich Schiller University, Institute of Applied Physics, Albert-Einstein-Str. 15, 07745 Jena, German
   \and  Institut für Halbleitertechnik, Technische Universität Braunschweig, Hans-Sommer-Str. 66, 38106 Braunschweig, Germany
   \and Laboratory for Emerging Nanometrology, Langer Kamp 6a-b, 38106 Braunschweig, Germany
   \and Fraunhofer-Institute for Applied Optics and Precision Engineering IOF, Albert-Einstein-Str. 7, 07745 Jena, Germany
   \and Physikalisch-Technische Bundesanstalt, Bundesallee 100, 38116 Braunschweig, Germany
   \and Ernst-Abbe-Hochschule Jena University of Applied Sciences, Carl-Zeiss-Promenade 2, 07745 Jena, Germany
   }

   \date{Received April 15, 2026}

 
  \abstract
{Coating thermal noise in high-reflectivity test-mass-mirrors is a major limitation for  observations between $\approx10-300$~Hz for the sensitivity of future gravitational-wave detectors. In particular, ET-Pathfinder requires mirror coatings that combine extremely high reflectance at 1.55~\textmu m with low thermal noise under cryogenic conditions. Conventional dielectric Bragg mirrors achieve high reflectance at the cost of increased coating thickness and mechanical dissipation, while metasurface-based mirrors can reduce coating-related noise but have not yet demonstrated comparable reflectance under realistic fabrication conditions.}
{We present a novel hybrid metasurface-Bragg mirror concept tailored to meet the requirements of ET-Pathfinder. The design combines a fabrication-tolerant, one-layer metasurface with an anti-resonant Fabry-Pérot spacer and a reduced dielectric Bragg stack. Fabrication effects, including geometric tolerances and line-edge roughness (LER), are explicitly taken into account.}
{The optical performance is evaluated using full-wave electromagnetic simulations. Fabrication robustness is analysed via a truncated Gaussian Monte Carlo approach, while the influence of LER is modelled as a systematic edge smoothing effect. The resulting reflectance distributions are used to determine the minimum number of Bragg layers required to meet system-level specifications. Thermal noise is calculated using a finite-element approach under ET-Pathfinder conditions.}
{The ideal metasurface design achieves reflectance values exceeding 99.999\%. Including the worst case assumptions, fabrication uncertainties and LER, the reflectance is limited to $\approx99.9\%$ at a 95\% yield level. The residual transmission can be compensated by a supporting Bragg stack with as few as seven layer pairs. For this configuration, the hybrid mirror achieves a total thermal displacement noise approximately one order of magnitude below the projected ET-Pathfinder coating noise budget.}
{These results demonstrate that fabrication-limited metasurface reflectance can be effectively compensated within a hybrid architecture, enabling a substantial reduction of coating thickness and thermal noise. The presented concept provides a practical pathway toward low-noise, high-reflectivity mirrors for next-generation gravitational-wave detectors.}


   \maketitle

\begin{figure*}[h!]
    \centering
    \includegraphics[width=1\linewidth]{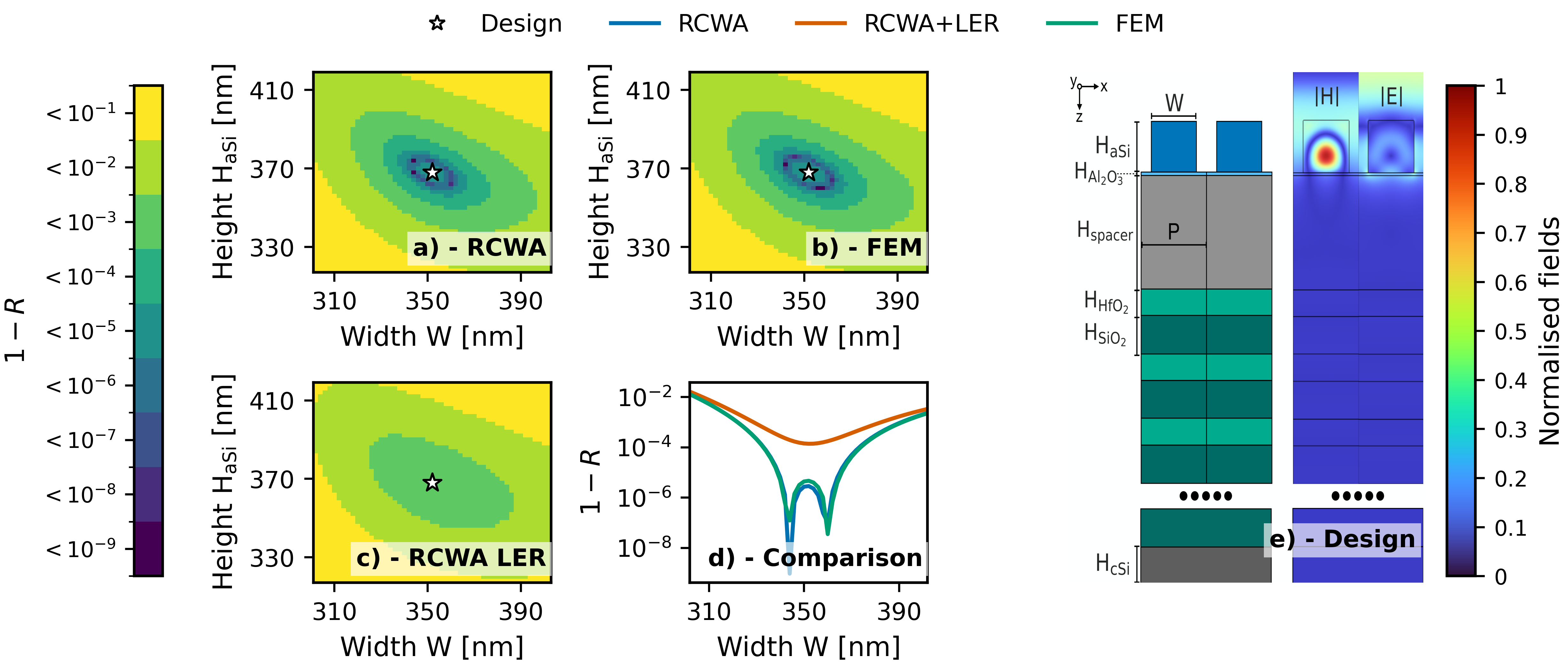}
    \caption{1 - reflection maps in the dependence of width W and height $\mathrm{H_{aSi}}$ ($\pm50$~nm at 1.55~µm) and the hybrid mirror design for constant period of 568~nm: a) RCWA simulation, b) FEM simulation, c) RCWA with line edge roughness (LER), d) comparison plot between RCWA without LER, RCWA with LER and FEM with only variation in width for the optimal parameters, e) design with normalised magnetic  $\lvert\mathrm{H}\rvert$ and electric field $\lvert\mathrm{E}\rvert$ showing the interaction of the TE-polarised light with the metasurface structure, the Fabry-Pérot spacer and the supporting Bragg-mirror stack.}
    \label{fig:reflection_map}
\end{figure*}

\section{Introduction}

Coating thermal noise is one of the key limitations to the sensitivity of present and future gravitational-wave detectors \citep{degallaix2019a}, particularly in the frequency range between $\approx$10--300~Hz in which compact-binary inspiral signals accumulate a large fraction of their signal-to-noise ratio (for more science cases see \citet{maggiore2020a}). Reducing coating thickness while maintaining ultra-high reflectance is a key challenge for ET-Pathfinder.

Metasurfaces \citep{kuznetsov2024a} offer a promising route to address this trade-off because they can generate strong optical resonances within a single sub-wavelength structured layer and thereby substantially reduce the effective optical thickness of the reflective element \citep{kroker2017a}. However, high-reflectance metasurfaces are generally sensitive to fabrication imperfections and is challenging to transfer to large-area optics with the robustness required for gravitational-wave instrumentation. A hybrid concept that combines a resonant metasurface with a reduced supporting Bragg stack is therefore attractive: the metasurface can provide the dominant reflectance contribution, while the Bragg mirror compensates the residual transmission needed to meet the system-level optical target. While \citet{dickmann2023a} separates both optical elements completely electro-magnetic \& mechanically with a Fabry-Pérot spacer in the cm-regime, our version presents the minimum thickness to decouple both elements electro-magnetically. The design is tailored to meet the requirements of ET-Pathfinder \citep{ETPathfinderReport2020} except for thickness and curvature as the test-platform of the Einstein Telescope. Its specify a 6-inch mirror diameter, operation at 1.55~\textmu m, a residual transmission of $1 - R \leq 10^{-5}$, and a thermal displacement noise below $10^{-18}\,\mathrm{m/\sqrt{Hz}}$ at 30~Hz under cryogenic conditions (10--20~K) for a beam radius of 2.2~mm \citep{Utina_2022}. 

Here we present a fabrication-aware hybrid metasurface-Bragg mirror concept where we present a worst case scenario based on our current clean-room process chain. We first analyse the ideal optical response of the design and then quantify two distinct fabrication-related limitations: the systematic reflectance reduction introduced by line-edge roughness \citep{siefke2018a, siefke2024a} and the statistical reflectance spread caused by  fabrication tolerances. These optical results are subsequently linked to the number of supporting Bragg pairs required to satisfy ET-Pathfinder-level reflectance and to the resulting thermal displacement noise \citep{harry2002} of the hybrid mirror. In this way, the letter connects fabrication realism, optical performance, and thermal-noise reduction within one consistent design framework. This framework allows also the adaption to different fabrication technologies and if applicable smaller tolerances.

\section{Hybrid Mirror Design}

The proposed mirror architecture combines a resonant metasurface with a reduced dielectric Bragg stack and an anti-resonant Fabry-Pérot spacer (Fig.~\ref{fig:reflection_map}e; parameters in Tab.~\ref{table:design}). The metasurface consists of a one-dimensional amorphous silicon grating supporting a Mie-type resonance for TE (s-)polarised light, characterised by a strongly localised magnetic field within the structure. This resonance enables near-unity reflectance while substantially reducing the effective optical thickness compared with conventional multilayer coatings. The hybrid concept addresses these constraints by assigning the dominant reflectance contribution to the metasurface, while a reduced Bragg stack compensates the remaining transmission required to meet the system-level optical target. 

The structure is implemented on a float-zone crystalline silicon substrate and comprises (from bottom to top) a Bragg stack of alternating amorphous $\mathrm{HfO_2}$ and $\mathrm{SiO_2}$ layers, an amorphous $\mathrm{SiO_2}$ spacer, an $\mathrm{Al_2O_3}$ etch-stop layer, and the structured amorphous silicon metasurface. The geometrical parameters are optimised using rigorous coupled-wave analysis (RCWA) \citep{moharam1981a, jin2020a} in combination with gradient-free optimisation \citep{nevergrad}. All used refractive indices for the deposited films are found in Tab.~\ref{table:matparam}. 

Figure~\ref{fig:reflection_map}a illustrates the ideal reflectance landscape of the metasurface as a function of structure height and width. In the absence of fabrication imperfections, reflectance values exceeding 99.999\% are obtained at 1.55~\textmu m over a width \& height window of approximately 20~nm. A comparison between RCWA and finite-element method (FEM) simulations (Fig.~\ref{fig:reflection_map}a,b,d) shows excellent agreement, with deviations below $1 - R \leq 10^{-6}$, confirming that numerical uncertainties are at least one order of magnitude  below the required reflectance threshold.

These ideal results assume perfectly sharp material interfaces. In practice, fabrication processes introduce line-edge roughness (LER). In order to include this into the simulation, a simplified model is used \citep{siefke2024a}, which leads to a sigmoidal smoothing of the refractive-index profile at the metasurface boundaries. To quantify this effect, we model LER by applying a Gaussian smoothing of the refractive-index profile with a full width at half maximum of 6.8~nm \citep{siefke2018a}, corresponding to the characteristic roughness of our lithography and etching processes. This model primarily captures the influence of high-spatial-frequency roughness components and therefore represents a worst-case estimate of the resulting optical degradation.

\begin{figure*}
    \centering
    \includegraphics[width=1\linewidth]{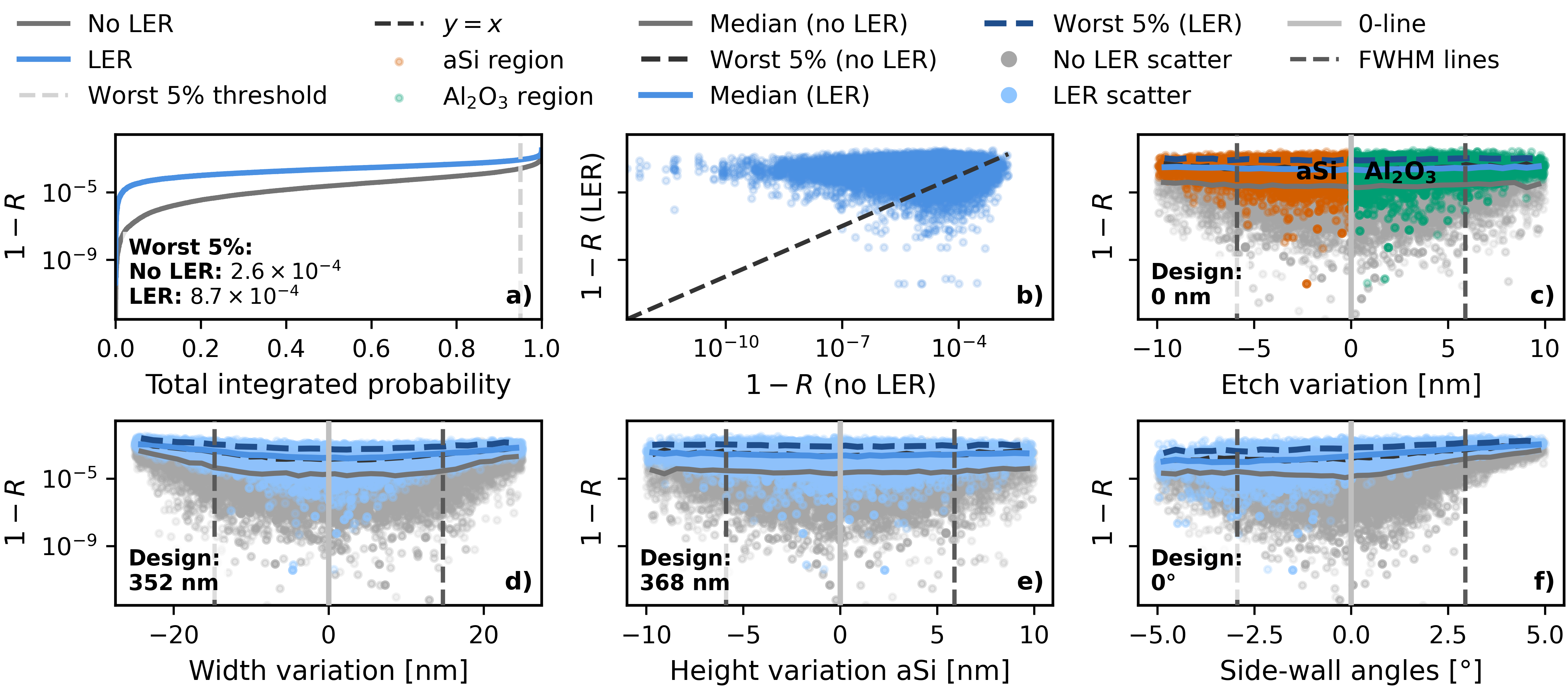}
    \caption{
Monte Carlo tolerance analysis of metasurface 1 - reflectance (better visibility) including fabrication uncertainties. Two data sets are shown: without line-edge roughness (LER, grey) and with LER (blue). a) Cumulative probability distribution of $1 - R$, illustrating the 95\% yield criterion. LER introduces a systematic shift towards higher residual transmission. b) Correlation between $1 - R$ with and without LER for each Monte Carlo realisation while the black dashed line represents equal $1 - R$ values. c–f) Sensitivity of $1 - R$ to individual fabrication parameters.
}
    \label{fig:mc}
\end{figure*}

Including this worst-case estimate for LER (Fig.~\ref{fig:reflection_map}c,d) the achievable reflectance is reduced to approximately 99.9\%. A comparison between the ideal and smoothed structures reveals two distinct regimes: below 99\% reflectance, the influence of LER is minor, whereas above 99\% it becomes the dominant limiting factor. In this high-reflectance regime, fabrication-induced interface smoothing rather than intrinsic design limitations determines the achievable performance.

The ET-Pathfinder-compatible reflectance under realistic fabrication conditions for a metasurface alone therefore is very challenging. To compensate for the fabrication aware reduction, a supporting Bragg stack is introduced. In combination with the anti-resonant Fabry–Pérot spacer, this hybrid configuration restores near-unity reflectance while maintaining a substantially reduced total coating thickness, which is essential for lowering the associated thermal-noise contributions.

\section{Metasurface Fabrication Uncertainties \& Reflection Probabilities}

After establishing the systematic reflectance limitation introduced by line-edge roughness (LER), we quantify the statistical fabrication yield of the metasurface under realistic process variations using a Monte Carlo tolerance analysis. While the LER model defines a process-dependent upper performance ceiling for the nominal design, the Monte Carlo analysis addresses a complementary question: how fabrication-induced parameter variations distribute the reflectance around this ceiling.

We model fabrication uncertainties using truncated Gaussian distributions based on realistic tolerances. Varied parameters include period ($\pm1$~nm), width ($\pm25$~nm), height ($\pm10$~nm), sidewall angle $\pm5^\circ$), and refractive indices ($\pm0.3$\%). For each realisation, the optical response is recalculated using RCWA, ensuring physically consistent geometries.

Figure~\ref{fig:mc} shows the resulting reflectance distributions and parameter sensitivities. In the absence of LER, a reflectance of $\geq99.97\%$ is achieved with 95\% probability. Including LER shifts this distribution to $\geq99.91\%$ at 95\% probability, consistent with the systematic reflectance reduction identified in the previous section. The Monte Carlo results therefore confirm that fabrication-induced parameter variations broaden the reflectance distribution, while LER defines the dominant upper limitation. The analysis further identifies the metasurface width and positive sidewall angle as the most critical fabrication parameters, directly affecting the resonance condition and therefore the reflectance. In contrast, other parameters such as period and refractive-index variations contribute less significantly.

It is important to note that both the LER treatment and the Monte Carlo sampling represent conservative assumptions. The presented 95\% yield values should therefore be interpreted as lower-bound performance estimates rather than as the most probable realised device performance.

These results demonstrate that the metasurface design remains robust under realistic fabrication conditions, while also establishing that the LER-induced reflectance ceiling must be compensated by a supporting Bragg stack in the hybrid configuration.

\begin{figure*}[h!]
    \centering
    \includegraphics[width=1\linewidth]{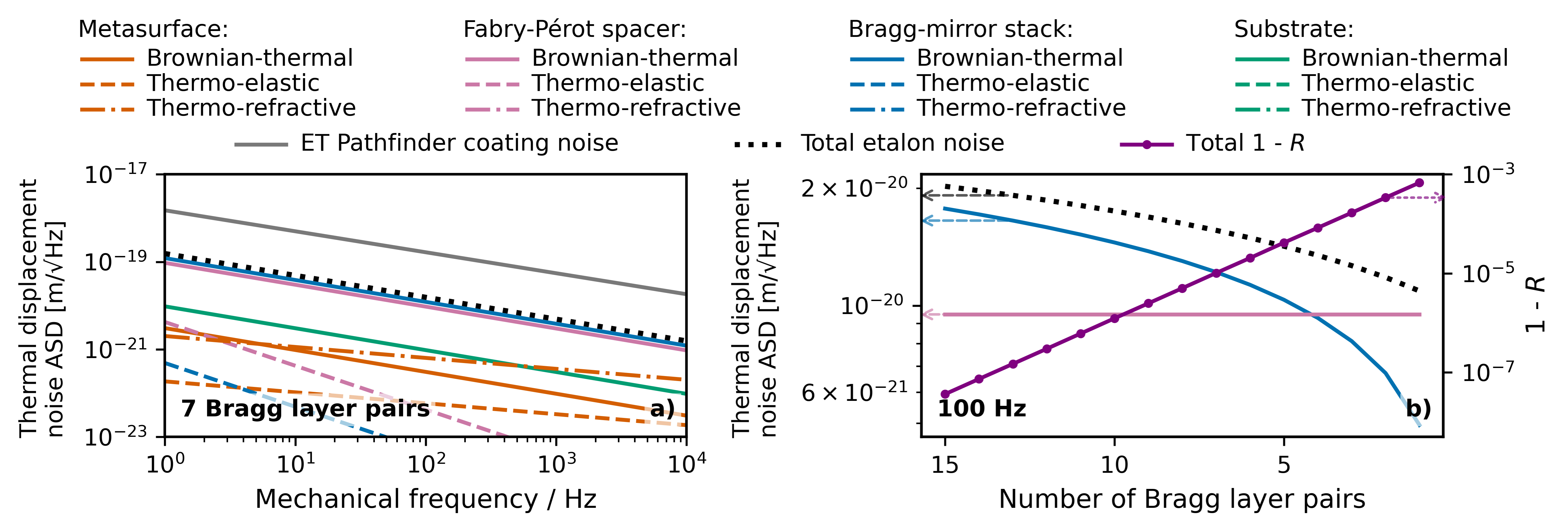}
    \caption{
Thermal displacement noise of the hybrid mirror under ET-Pathfinder conditions (info: $\mathrm{Al_2O_3}$ is included in the Fabry-Pérot spacer). a) Contributions of Brownian, thermo-elastic, and thermo-refractive  noise from all components, compared to the projected ET-Pathfinder coating noise budget. b) Total thermal noise and $1-R$ as a function of the number of Bragg layer pairs.
}
    \label{fig:noise_comaprison}
\end{figure*}

\section{Supporting Bragg Layers \& Thermal Noise Analysis}

While the ideal metasurface achieves near-unity reflectance, the inclusion of line-edge roughness (LER) and fabrication-induced parameter variations reduces the achievable reflectance to approximately $R \approx 99.9\%$ at a 95\% yield level. The remaining transmission must therefore be compensated by a supporting Bragg stack to meet the ET-Pathfinder requirement of $1 - R \leq 10^{-5}$. To quantify the resulting thermal-noise performance, we evaluate the thermal displacement noise of the hybrid mirror under ET-Pathfinder conditions using the material parameters listed in Tab.~\ref{table:matparam}. The analysis is based on the nominal hybrid design, where the metasurface provides the dominant reflectance contribution and the Bragg stack compensates the residual transmission. 

Figure~\ref{fig:noise_comaprison}a shows the individual contributions of Brownian, thermoelastic (TE), and thermorefractive (TR) noise for all components of the hybrid mirror. The projected coating thermal displacement noise for ET-Pathfinder \citep{Utina_2022} is included for comparison. The total thermal noise is calculated as the incoherent sum of all contributions following \citet{dickmann2018b}. All layers contribute to Brownian noise, making Bragg-layer reduction critical.

Figure~\ref{fig:noise_comaprison}b shows the total thermal noise and residual transmission as a function of the number of Bragg layer pairs. Starting from the fabrication-limited reflectance of the metasurface, the Bragg stack dependence of noise on number of layer pairs is incrementally reduced to identify the minimum number of layers required to meet the ET-Pathfinder reflectance specification. We find that a configuration with seven Bragg layer pairs is sufficient to comply with the design requirements while maintaining a significantly reduced coating thickness. For this configuration, the total thermal displacement noise is approximately one order of magnitude below the projected coating noise budget of ET-Pathfinder. This reduction is primarily enabled by transferring the dominant reflectance function from the Bragg stack to the metasurface, thereby minimising the mechanically lossy coating volume.

For very small numbers of Bragg layers, the total noise becomes dominated by the spacer layer due to the comparatively high mechanical loss of $\mathrm{SiO_2}$ at cryogenic temperatures ($\sim5\times10^{-4}$; \citet{franc2009mirrorthermalnoiselaser}). While this contribution could be reduced by employing lower-loss spacer materials such as sapphire \citep{Uchiyama1999}, such a change would require a re-optimisation of the optical design.

\section{Conclusions}

We presented a fabrication-tolerant hybrid metasurface–Bragg mirror design meeting ET-Pathfinder requirements. Including fabrication uncertainties and line-edge roughness (LER), the metasurface reflectance is limited to $\approx$99.9\% at 95\% yield. A minimal supporting Bragg stack of seven layer pairs compensates the residual transmission, substantially reducing coating thickness and mechanical loss.

The hybrid mirror achieves a thermal displacement noise approximately one order of magnitude below the projected ET-Pathfinder coating noise budget. Performance is limited by LER rather than intrinsic design constraints, with width and sidewall angle identified as the dominant fabrication parameters.

By directly linking fabrication-limited reflectance to thermal-noise reduction, this approach enables ultra-low-noise mirrors with minimal multilayer coatings. The concept establishes a scalable route toward high-reflectivity, low-noise optics for next-generation gravitational-wave detectors, including the Einstein Telescope, and other precision optical systems such as optical clocks and ultra-stable cavities.

\begin{acknowledgements}

 Part of this work was supported by the \textit{German Federal Ministry of Research, Technology and Space} (\emph{BMFTR}), chapter 4, title 89450, fiscal year 2024, project: „Verbundprojekt 05A2023 - 3G-GWD: Gravitationswellenteleskope der dritten Generation. Teilprojekt 12.“, funding ID number 05A24SJ1. The authors acknowledge partial support by the Deutsche Forschungsgemeinschaft (DFG, German Research Foundation) under Germany’s Excellence Strategy – EXC 2123/2 QuantumFrontiers – 390837967 and support by the German Federal Ministry of Research, Technology and Space (BMFTR) under grant number 05A2023. 

\end{acknowledgements}

%

\bibliographystyle{aa}
\bibliography{20250103_MetaMirrors}

@article{dickmann2018b,
  title = {Highly Reflective Low-Noise Etalon-Based Meta-Mirror},
  author = {Dickmann, Johannes and Kroker, Stefanie},
  year = 2018,
  month = oct,
  journal = {Physical Review D},
  volume = {98},
  number = {8},
  pages = {082003},
  doi = {10.1103/PhysRevD.98.082003},
  urldate = {2025-01-03}
}

@article{dickmann2023a,
  title = {Experimental Realization of a 12,000-Finesse Laser Cavity Based on a Low-Noise Microstructured Mirror},
  author = {Dickmann, Johannes and Sauer, Steffen and Meyer, Jan and Gaedtke, Mika and Siefke, Thomas and Br{\"u}ckner, Uwe and Plentz, Jonathan and Kroker, Stefanie},
  year = 2023,
  month = jan,
  journal = {Communications Physics},
  volume = {6},
  number = {1},
  pages = {1--5},
  issn = {2399-3650},
  doi = {10.1038/s42005-023-01131-1},
  urldate = {2025-01-03}
}

@article{jin2020a,
  title = {Inverse {{Design}} of {{Lightweight Broadband Reflector}} for {{Relativistic Lightsail Propulsion}}},
  author = {Jin, Weiliang and Li, Wei and Orenstein, Meir and Fan, Shanhui},
  year = 2020,
  month = sep,
  journal = {ACS Photonics},
  volume = {7},
  number = {9},
  pages = {2350--2355},
  doi = {10.1021/acsphotonics.0c00768},
  urldate = {2025-01-09}
}

@article{kroker2017a,
  title = {Brownian Thermal Noise in Functional Optical Surfaces},
  author = {Kroker, S. and Dickmann, J. and Rojas Hurtado, C. B. and Heinert, D. and Nawrodt, R. and Levin, Y. and Vyatchanin, S. P.},
  year = 2017,
  month = jul,
  journal = {Physical Review D},
  volume = {96},
  number = {2},
  pages = {022002},
  doi = {10.1103/PhysRevD.96.022002},
  urldate = {2025-01-03}
}

@article{kuznetsov2024a,
  title = {Roadmap for {{Optical Metasurfaces}}},
  author = {Kuznetsov, Arseniy I. and Brongersma, Mark L. and Yao, Jin and Chen, Mu Ku and Levy, Uriel and Tsai, Din Ping and Zheludev, Nikolay I. and Faraon, Andrei and Arbabi, Amir and Yu, Nanfang and Chanda, Debashis and Crozier, Kenneth B. and Kildishev, Alexander V. and Wang, Hao and Yang, Joel K. W. and Valentine, Jason G. and Genevet, Patrice and Fan, Jonathan A. and Miller, Owen D. and Majumdar, Arka and Fr{\"o}ch, Johannes E. and Brady, David and Heide, Felix and Veeraraghavan, Ashok and Engheta, Nader and Al{\`u}, Andrea and Polman, Albert and Atwater, Harry A. and Thureja, Prachi and {Paniagua-Dominguez}, Ramon and Ha, Son Tung and Barreda, Angela I. and Schuller, Jon A. and Staude, Isabelle and Grinblat, Gustavo and Kivshar, Yuri and Peana, Samuel and Yelin, Susanne F. and Senichev, Alexander and Shalaev, Vladimir M. and Saha, Soham and Boltasseva, Alexandra and Rho, Junsuk and Oh, Dong Kyo and Kim, Joohoon and Park, Junghyun and Devlin, Robert and Pala, Ragip A.},
  year = 2024,
  month = mar,
  journal = {ACS Photonics},
  volume = {11},
  number = {3},
  pages = {816--865},
  doi = {10.1021/acsphotonics.3c00457},
  urldate = {2025-01-06}
}

@article{maggiore2020a,
  title = {Science Case for the {{Einstein}} Telescope},
  author = {Maggiore, Michele and Broeck, Chris Van Den and Bartolo, Nicola and Belgacem, Enis and Bertacca, Daniele and Bizouard, Marie Anne and Branchesi, Marica and Clesse, Sebastien and Foffa, Stefano and {Garc{\'i}a-Bellido}, Juan and Grimm, Stefan and Harms, Jan and Hinderer, Tanja and Matarrese, Sabino and Palomba, Cristiano and Peloso, Marco and Ricciardone, Angelo and Sakellariadou, Mairi},
  year = 2020,
  month = mar,
  journal = {Journal of Cosmology and Astroparticle Physics},
  volume = {2020},
  number = {3},
  pages = {050},
  issn = {1475-7516},
  doi = {10.1088/1475-7516/2020/03/050},
  urldate = {2025-01-06}
}

@article{moharam1981a,
  title = {Rigorous Coupled-Wave Analysis of Planar-Grating Diffraction},
  author = {Moharam, M. G. and Gaylord, T. K.},
  year = 1981,
  month = jul,
  journal = {JOSA},
  volume = {71},
  number = {7},
  pages = {811--818},
  doi = {10.1364/JOSA.71.000811},
  urldate = {2025-01-07}
}

@article{siefke2018a,
  title = {Line-Edge Roughness as a Challenge for High-Performance Wire Grid Polarizers in the Far Ultraviolet and Beyond},
  author = {Siefke, Thomas and Heusinger, Martin and Hurtado, Carol B. Rojas and Dickmann, Johannes and Zeitner, Uwe and T{\"u}nnermann, Andreas and Kroker, Stefanie},
  year = 2018,
  month = jul,
  journal = {Optics Express},
  volume = {26},
  number = {15},
  pages = {19534--19547},
  issn = {1094-4087},
  doi = {10.1364/OE.26.019534},
  urldate = {2025-01-03}
}

@misc{nevergrad,
    author = {J. Rapin and O. Teytaud},
    title = {{Nevergrad - A gradient-free optimization platform}},
    year = {2018},
    publisher = {GitHub},
    journal = {GitHub repository},
    howpublished = {\url{https://GitHub.com/FacebookResearch/Nevergrad}},
}

@article{harry2002,
  title = {Thermal Noise in Interferometric Gravitational Wave Detectors Due to Dielectric Optical Coatings},
  author = {Harry, Gregory M. and Gretarsson, Andri M. and Saulson, Peter R. and Kittelberger, Scott E. and Penn, Steven D. and Startin, William J. and Rowan, Sheila and Fejer, Martin M. and Crooks, D. R. M. and Cagnoli, Gianpietro and Hough, Jim and Nakagawa, Norio},
  year = 2002,
  month = feb,
  journal = {Classical and Quantum Gravity},
  volume = {19},
  number = {5},
  pages = {897},
  issn = {0264-9381},
  doi = {10.1088/0264-9381/19/5/305},
  url = {https://doi.org/10.1088/0264-9381/19/5/305},
  urldate = {2026-03-25},
  langid = {english}
}

@article{Utina_2022,
doi = {10.1088/1361-6382/ac8fdb},
url = {https://doi.org/10.1088/1361-6382/ac8fdb},
year = {2022},
month = {sep},
publisher = {IOP Publishing},
volume = {39},
number = {21},
pages = {215008},
author = {Utina, A and Amato, A and Arends, J and Arina, C and de Baar, M and Baars, M and Baer, P and van Bakel, N and Beaumont, W and Bertolini, A and van Beuzekom, M and Biersteker, S and Binetti, A and ter Brake, H J M and Bruno, G and Bryant, J and Bulten, H J and Busch, L and Cebeci, P and Collette, C and Cooper, S and Cornelissen, R and Cuijpers, P and van Dael, M and Danilishin, S and Diksha, D and van Doesburg, S and Doets, M and Elsinga, R and Erends, V and van Erps, J and Freise, A and Frenaij, H and Garcia, R and Giesberts, M and Grohmann, S and Van Haevermaet, H and Heijnen, S and van Heijningen, J V and Hennes, E and Hennig, J-S and Hennig, M and Hertog, T and Hild, S and Hoffmann, H-D and Hoft, G and Hopman, M and Hoyland, D and Iandolo, G A and Ietswaard, C and Jamshidi, R and Jansweijer, P and Jones, A and Jones, P and Knust, N and Koekoek, G and Koroveshi, X and Kortekaas, T and Koushik, A N and Kraan, M and van de Kraats, M and Kranzhoff, S L and Kuijer, P and Kukkadapu, K A and Lam, K and Letendre, N and Li, P and Limburg, R and Linde, F and Locquet, J-P and Loosen, P and Lueck, H and Martínez, M and Masserot, A and Meylahn, F and Molenaar, M and Mow-Lowry, C and Mundet, J and Munneke, B and van Nieuwland, L and Pacaud, E and Pascucci, D and Petit, S and Van Ranst, Z and Raskin, G and Recaman, P M and van Remortel, N and Rolland, L and de Roo, L and Roose, E and Rosier, J C and Ryckbosch, D and Schouteden, K and Sevrin, A and Sider, A and Singha, A and Spagnuolo, V and Stahl, A and Steinlechner, J and Steinlechner, S and Swinkels, B and Szilasi, N and Tacca, M and Thienpont, H and Vecchio, A and Verkooijen, H and Vermeer, C H and Vervaeke, M and Visser, G and Walet, R and Werneke, P and Westhofen, C and Willke, B and Xhahi, A and Zhang, T},
title = {ETpathfinder: a cryogenic testbed for interferometric gravitational-wave detectors},
journal = {Classical and Quantum Gravity}
}

@ARTICLE{Uchiyama1999,
       author = {{Uchiyama}, T. and {Tomaru}, T. and {Tobar}, M.~E. and {Tatsumi}, D. and {Miyoki}, S. and {Ohashi}, M. and {Kuroda}, K. and {Suzuki}, T. and {Sato}, N. and {Haruyama}, T. and {Yamamoto}, A. and {Shintomi}, T.},
        title = "{Mechanical quality factor of a cryogenic sapphire test mass for gravitational wave detectors}",
      journal = {Physics Letters A},
         year = 1999,
        month = oct,
       volume = {261},
       number = {1-2},
        pages = {5-11},
          doi = {10.1016/S0375-9601(99)00563-0},
       adsurl = {https://ui.adsabs.harvard.edu/abs/1999PhLA..261....5U}
}

@misc{franc2009mirrorthermalnoiselaser,
      title={Mirror thermal noise in laser interferometer gravitational wave detectors operating at room and cryogenic temperature},
      author={Janyce Franc and Nazario Morgado and Raffaele Flaminio and Ronny Nawrodt and Iain Martin and Liam Cunningham and Alan Cumming and Sheila Rowan and James Hough},
      year={2009},
      eprint={0912.0107},
      archivePrefix={arXiv},
      journal = {ArXiv e-prints},
      primaryClass={gr-qc},
      url={https://arxiv.org/abs/0912.0107}, 
}

@article{PhysRevD.92.062001,
  title = {Ion-beam sputtered amorphous silicon films for cryogenic precision measurement systems},
  author = {Murray, Peter G. and Martin, Iain W. and Craig, Kieran and Hough, James and Robie, Raymond and Rowan, Sheila and Abernathy, Matt R. and Pershing, Teal and Penn, Steven},
  journal = {Phys. Rev. D},
  volume = {92},
  issue = {6},
  pages = {062001},
  numpages = {11},
  year = {2015},
  month = {Sep},
  publisher = {American Physical Society},
  doi = {10.1103/PhysRevD.92.062001},
  url = {https://link.aps.org/doi/10.1103/PhysRevD.92.062001}
}

@article{Alcala_2002,
doi = {10.1088/0957-4484/13/4/302},
url = {https://doi.org/10.1088/0957-4484/13/4/302},
year = {2002},
month = {may},
publisher = {},
volume = {13},
number = {4},
pages = {451},
author = {G Alcalá and P Skeldon and G E Thompson and A B Mann and H Habazaki and K Shimizu},
title = {Mechanical properties of amorphous anodic alumina and tantala films using nanoindentation},
journal = {Nanotechnology}
}

@book{Ventura2014ThermalProperties,
  author    = {Guglielmo Ventura and Mauro Perfetti},
  title     = {Thermal Properties of Solids at Room and Cryogenic Temperatures},
  year      = {2014},
  publisher = {Springer},
  address   = {Dordrecht},
  series    = {International Cryogenics Monograph Series},
  doi       = {10.1007/978-94-017-8969-1},
  isbn      = {978-94-017-8969-1}
}

@article{PhysRevLett.96.055902,
  title = {Thermal Conductivity and Specific Heat of Thin-Film Amorphous Silicon},
  author = {Zink, B. L. and Pietri, R. and Hellman, F.},
  journal = {Phys. Rev. Lett.},
  volume = {96},
  issue = {5},
  pages = {055902},
  numpages = {4},
  year = {2006},
  month = {Feb},
  publisher = {American Physical Society},
  doi = {10.1103/PhysRevLett.96.055902},
  url = {https://link.aps.org/doi/10.1103/PhysRevLett.96.055902}
}

@article{siefke2024a,
  title = {Influence of Line Edge Roughness in Optical Critical Dimension Metrology},
  author = {{Siefke, Thomas}},
  year = 2024,
  journal = {Epj Web of Conferences},
  volume = {309},
  pages = {02011},
  doi = {10.1051/epjconf/202430902011},
  url = {https://doi.org/10.1051/epjconf/202430902011}
}

@article{Franta:15,
author = {Daniel Franta and David Ne\v{c}as and Ivan Ohl\'{i}dal},
journal = {Appl. Opt.},
number = {31},
pages = {9108--9119},
publisher = {Optica Publishing Group},
title = {Universal dispersion model for characterization of optical thin films over a wide spectral range: application to hafnia},
volume = {54},
month = {Nov},
year = {2015},
url = {https://opg.optica.org/ao/abstract.cfm?URI=ao-54-31-9108},
doi = {10.1364/AO.54.009108},
}

@book{Palik1985OpticalConstants,
  editor    = {Edward D. Palik},
  title     = {Handbook of Optical Constants of Solids},
  publisher = {Academic Press},
  year      = {1985},
  isbn      = {978-0-12-544415-6}
}

@techreport{ETPathfinderReport2020,
    author = {{The ETpathfinder Team}},
    title = {ETpathfinder
DESIGN REPORT},
    institution = {Nikhef, Maastricht University, University of Antwerp, Ghent University,
Katholieke Universiteit Leuven, Université Catholique de Louvain, Hasselt
University, Vrije Universiteit Brussel, Fraunhofer Institute for Laser Technology,
RWTH Aachen, University of Twente, Eindhoven University of Technology,
Liege Université, VITO, TNO},
    year = {2020}
}

@article{degallaix2019a,
author = {J\'{e}r\^{o}me Degallaix and Christophe Michel and Benoit Sassolas and Annalisa Allocca and Gianpetro Cagnoli and Laurent Balzarini and Vincent Dolique and Raffaele Flaminio and Dani\`{e}le Forest and Massimo Granata and Bernard Lagrange and Nicolas Straniero and Julien Teillon and Laurent Pinard},
journal = {J. Opt. Soc. Am. A},
number = {11},
pages = {C85--C94},
publisher = {Optica Publishing Group},
title = {Large and extremely low loss: the unique challenges of gravitational wave mirrors},
volume = {36},
month = {Nov},
year = {2019},
url = {https://opg.optica.org/josaa/abstract.cfm?URI=josaa-36-11-C85},
doi = {10.1364/JOSAA.36.000C85},
}

\begin{appendix}





\clearpage


\onecolumn

\section{Tables}


We present here the values used for our simulations. 

\begin{table}[ht!]
\caption{\label{table:design}Design parameters for hybrid metasurface-Bragg mirror concept}
\centering
\begin{tabular}{lll}
\hline\hline
Layer&Material&Parameters\\
\hline
Metasurface&aSi&Period\\
& &$\mathrm{P}=$ 568~nm,\\
& & Height \\
& & $\mathrm{H}_\mathrm{aSi}=$ 368~nm,\\
& & Width\\
& & $\mathrm{W}=$ 352~nm\\
\hline
Etch-stop layer&a$\mathrm{Al_2O_3}$&Height\\ 
& & $\mathrm{H}_\mathrm{Al_2O_3}=$ 20~nm\\ 
\hline
Fabry-Pérot spacer & a$\mathrm{SiO_2}$ & Height \\ 
&   &  $\mathrm{H}_\mathrm{spacer}=$ 805~nm\\ 
\hline
Bragg-mirror-stack & a$\mathrm{HfO_2}$ / & Layers  \\
& a$\mathrm{SiO_2}$ &   7\\
\hline
Substrate & cSi & Height \\
 &  & $\mathrm{H}_\mathrm{cSi}=$ 1~mm\\
\hline
\end{tabular}
\end{table}

\begin{table*}[h!]
\caption{Material parameters used for calculations. }
\label{table:matparam} 
\centering
\begin{tabular}{lcccc}
\hline\hline             
Parameter & Silicon &$\mathrm{SiO_2}$ & $\mathrm{HfO_2}$ & $\mathrm{Al_2O_3}$\\
\hline
n & 3.774 & 1.444\tablefootmark{a} & 2.052\tablefootmark{b} & 1.619\\
$\mathrm{Y\,[GPa]}$ & 130 & 72 & 380 & 120\tablefootmark{c}\\
$\mathrm{\nu}$ & 0.22 & 0.17 & 0.2 & 0.22\\
$\mathrm{\rho\,[kg/m^3]}$ & 2331 & 2200 & 8000 & 3700\\
$\mathrm{\Phi\,[rad]}$ & $1\cdot10^{-5}$\tablefootmark{d} & $5\cdot10^{-4}$ & $2\cdot10^{-4}$ & $2.4\cdot10^{-4}$\\
$\mathrm{\kappa\,[W/(m\cdot K)]}$ & 0.2\tablefootmark{e} & 0.7 & 1.2 & 5\\
$\mathrm{C\,[J/(kg\cdot K)]}$ & 3.5\tablefootmark{f}& 38\tablefootmark{g} & 16.7 & 2.4\tablefootmark{h}\\
$\mathrm{\alpha\,[1/K]}$ & $4.85\cdot10^{-10}$ & $-0.25\cdot10^{-6}$ & $3.8\cdot10^{-6}$ & $0.6\cdot10^{-6}$\\
$\mathrm{\beta\,[1/K]}$ & $5.8\cdot10^{-6}$ & $1.01\cdot10^{-6}$  & $1\cdot10^{-6}$ & $1.3\cdot10^{-5}$\\

\hline
\end{tabular}

\tablefoot{The refractive indices of Silicon and $\mathrm{Al_2O_3}$ were measured by ellipsometry. All other values were taken from \cite{franc2009mirrorthermalnoiselaser} unless explicitly stated.\\
\tablefoottext{a}{Refractive index value from \cite{Palik1985OpticalConstants},}
\tablefoottext{b}{refractive index value from \cite{Franta:15},}
\tablefoottext{c}{Young's modulus value from \cite{Alcala_2002},}
\tablefoottext{d}{loss angle of the coating value from \cite{PhysRevD.92.062001},}
\tablefoottext{e}{specific thermal conductivity value from \cite{PhysRevLett.96.055902},}
\tablefoottext{f}{specific thermal capacity value from \cite{PhysRevLett.96.055902},}
\tablefoottext{g}{specific thermal capacity value from \cite{Ventura2014ThermalProperties},}
\tablefoottext{h}{specific thermal capacity value from \cite{Ventura2014ThermalProperties}.}
}

\end{table*}

\end{appendix}
\end{document}